\documentclass[twocolumn]{svjour3}       %
\usepackage{ifpdf}
\usepackage{graphicx,amssymb}
\usepackage{amsbsy}
\usepackage{amsmath} 
\usepackage{mathrsfs}
\usepackage{color}%

\newcommand{\sign}{\,{\rm sign}}

\newcommand{\eps}{\varepsilon}

\newcommand{\am}{{\rm \,am}}
\newcommand{\dn}{{\rm \,dn}}
\newcommand{\sn}{{\rm \,sn}}
\newcommand{\cn}{{\rm \,cn}}

\renewcommand{\d}{{\rm\,d}}

\begin{document}\sloppy
\title{Homoclinic, Subharmonic, and Superharmonic Bifurcations for a Pendulum with Periodically Varying Length}

\author{Anton~O.~Belyakov \and Alexander~P.~Seyranian
}

\institute{A.O. Belyakov \at
              Institute of Mechanics,
              Lomonosov Moscow State  University,
              Michurinsky pr. 1, 119192 Moscow, Russia\\
              \email{belyakov@imec.msu.ru}     \\
              \emph{Present address:} of A.O. Belyakov  \at
              ORCOS, Institute of Mathematical Methods in Economics, Vienna
              University of Technology\\Argentinierstrasse 8/105-4, A-1040 Vienna,
              Austria\\
              Tel.: +43 1 58801-105473\\
              Fax: +43 1 58801-9105473\\
              \email{anton.belyakov@tuwien.ac.at}
           \and
              A.P. Seyranian \at
              Institute of Mechanics,
              Lomonosov Moscow State  University,
              Michurinsky pr. 1, 119192 Moscow, Russia\\
              Tel.: (7495) 939 2039\\
              Fax: (7495) 939 0165\\
              \email{seyran@imec.msu.ru}
}

\date{Received: date / Accepted: date}

\maketitle

\begin{abstract}
Dynamic behavior of a weightless rod with a point mass sliding
along the rod axis according to periodic law is studied. This is
the simplest model of child's swing. Melnikov's analysis is
carried out to find bifurcations of homoclinic, subharmonic
oscillatory, and subharmonic rotational orbits. For the analysis
of superharmonic rotational orbits the averaging method is used
and stability of obtained approximate solution is checked.
The analytical results are compared with numerical simulation
results. 
\end{abstract}

\section{Introduction}
Oscillations of a pendulum with periodically varying length (PPVL)
is one of the classical problems in mechanics, see
\cite{Kauderer,BogMit,Pan,Mag,Seyran_swing,Seyran_Belyak,Zevin,BelSeyLuo,BelSey2012,Markeev,Markeev2012}.
We represent PPVL as a weightless rod with a point mass sliding
along the rod axis according to periodic law. This is also a
simple model of child's swing. In works
\cite{Kauderer,BogMit,Pan,Mag,Seyran_swing,Seyran_Belyak,Zevin,BelSeyLuo,BelSey2012}
small oscillations of PPVL were studied. In
\cite{Seyran_swing,BelSeyLuo,BelSey2012} the instability domains
of the vertical position were found both analytically and
numerically. Regular rotations and chaotic regimes were also
investigated, see \cite{Seyran_Belyak,BelSeyLuo,BelSey2012}.

In the literature oscillatory and rotational approximate solutions
for PPVL were obtained with quasi-linear approach, where
nonlinearity was assumed to be small as well as the excitation
amplitude. The only exception is \cite{Markeev,Markeev2012},
where exact stable uniform rotations were found in the case of zero damping
and harmonic excitation with special amplitude and phase.
But these uniform rotations occur only for very restrictive relation of parameters.
On the other hand, quasi-linear approach requires taking high order approximations for
substantially nonlinear regimes, such as regular rotations with frequencies higher than the frequency of excitation, see e.g. \cite{BelSey2012}. That makes approximate expressions cumbersome and analysis difficult, while the error of approximation can still be high, because the smallness assumption of both excitation and nonlinearity might not be satisfied for any existing rotations.

In order to resolve this issue in the present paper we analyze boundaries in the parameter space
for oscillatory, rotational, and more complex
(rotational-oscillatory, chaotic) regimes of {PPVL} assuming
arbitrary (not small) nonlinearity.

On one hand, the methods used for studying other parametrically
excited pendula, such as a pendulum with vibrating pivot
\cite{Koch_Leven,MannKoplow2006,XuWiercigroch2007}, can also be
applied to PPVL. In particular, we apply Melnikov analysis in a similar way to \cite{Koch_Leven}, that is applicable when excitation and damping are small.
We also compare the obtained boundaries with the results of numerical simulations.

On the other hand, our paper contains a methodological novelty since the
rotational regimes are studied without the assumption that excitation is small.
Instead of this, to apply the method of averaging, we
assume that the frequency of excitation is large. In that case
the unperturbed system preserves angular momentum, which is
a distinctive feature of PPVL. As a result the first order approximation
happens to be enough to find approximate solutions of different angular velocities.
We also study the stability of these solutions, find their existence domains in parameter space,
and compare them with the results of numerical simulations.

Numerical simulations were made in \cite{BelSeyLuo,BelSey2012} for different amplitudes and frequencies of excitation, starting from the same initial conditions. After sufficiently high time of simulation the solutions converged to regular regimes (equilibrium, oscillations, rotations) or remained ``chaotic''. Thus we have the points in parameter space, where corresponding regimes exist, for comparison with the analytically obtained approximations of the corresponding existence domains.

The paper is organized as follows. In Section \ref{s:main} main
equations of motion of a pendulum with variable length are derived
and given in non-dimensional form. In Section \ref{s:melnikov} we
use Melnikov's analysis
\cite{Melnikov,Guk,SzemplinskaStupnicka1995,Kuznetsov} to find
bifurcations of homoclinic, oscillatory and subharmonic rotational
orbits of PPVL.
 Melnikov's functions are obtained and
compared with the results of numerical simulation. In
Section~\ref{s:avg} we find bifurcations of superharmonic
rotational orbits with the use of the averaging method and compare
them with the numerical simulation results. In Section~\ref{s:avg}
we find the domains of existence of the rotational solutions under
assumption that the unperturbed system preserves the angular
momentum (rather than Hamiltonian as in the Melnikov's analysis)
and allows for non-small excitation amplitude.

\section{Main relations}\label{s:main}
Equation for motion of the PPVL is derived with the use of angular
momentum alteration theorem and taking into account linear damping
forces, see \cite{Seyran_swing,Seyran_Belyak,BelSeyLuo}
\begin{equation}\label{eq_swing}
    \frac{\d}{\d t}\left(m l^2 \frac{\d\theta}{\d t}\right)+\gamma l^2 \frac{\d\theta}{\d t} + m g l
    \sin(\theta) = 0,
\end{equation}
where $m$ is the mass, $l$ is the length, $\theta$ is the angle of
the pendulum deviation from the vertical position, $\gamma$ is the
damping coefficient, and $g$ is the acceleration due to gravity.

It is assumed that the length of the pendulum changes according to
a periodic law
\begin{equation}l = l_0 + a \varphi(\Omega t),\end{equation}
where $l_0$ is the mean pendulum length, $a$ and $\Omega$ are the
amplitude and frequency of the excitation, $\varphi(\tau)$ is a
smooth zero mean periodic function with period $2\pi$.

We introduce new time $\tau=\Omega t$ and three dimensionless
parameters
\begin{equation}\label{eq_param}\eps = \frac{a}{l_0},\quad \omega = \frac{\Omega_0}{\Omega},\quad \beta
= \frac{\gamma}{ m\Omega_0},
\end{equation}
where $\Omega_0 = \sqrt{\frac{g}{l_0}}$ is the eigenfrequency of
the pendulum with constant length $l = l_0$ and zero damping. In
this notations equation (\ref{eq_swing}) takes the form
\begin{equation}\label{eq:swing2}
\begin{split}
    \left(\left(1+\eps\varphi(\tau)\right)^2
    \dot{\theta}\right)\dot{} &
    +\beta\omega \left(1+\eps\varphi(\tau)\right)^2 \dot{\theta}\\
    & + \left(1+\eps\varphi(\tau)\right)\omega^2 \sin(\theta) = 0,
\end{split}
\end{equation} where the upper dot denotes
differentiation with respect to new time $\tau$.

\section{Melnikov's method: Perturbation of a Hamiltonian
System}\label{s:melnikov} Assuming $1+\eps\varphi(\tau)>0$,
equation (\ref{eq:swing2}) can be written in the following form
\begin{equation}\label{eq_theta}
    \ddot \theta + \left(\frac{2\eps \dot\varphi(\tau)}{1+\eps\varphi(\tau)} + \beta\omega\right)\dot{\theta}
    + \frac{\omega^2 \sin(\theta)}{1+\eps\varphi(\tau)} = 0.
\end{equation}
Coefficients of nonlinear equation (\ref{eq_theta}) explicitly
depend on the periodic function $\varphi(\tau)$ and  three
independent dimensionless parameters: the relative excitation
amplitude $\eps$, the damping $\beta$, and the inverse relative
frequency of excitation $\omega$.

Let us assume that parameters of excitation amplitude $\eps$ and
damping $\beta$ are small of the same order, $\eps\sim\beta \ll
1$. Thus, we can say that dynamics of PPVL is described by the
perturbed Hamiltonian system
\begin{eqnarray}
\dot{\theta} & = & \frac{\partial H}{\partial v},\label{eq:Hpert_theta}\\
\dot{v} & = & -\frac{\partial H}{\partial \theta} +
g_1(\theta,v,\tau)+o(\eps),\label{eq:Hpert_v}
\end{eqnarray}
where the perturbation function
\begin{equation}\label{eq:g1}
g_1(\theta,v,\tau) = \left(2\eps\sin(\tau)-\beta\omega\right)v +
\eps\omega^2\cos(\tau)\sin(\theta)
\end{equation}
is of the first order of smallness, i.e. $g_1(\theta,v,\tau)
=O(\eps)$. The following function
\begin{equation}\label{eq:H}
    H = \frac{v^2}{2} - \omega^2 \cos(\theta)
\end{equation}
is the Hamiltonian of the unperturbed system
\begin{eqnarray}
\dot{\theta} & = & v,\label{eq:Hunpert_theta}\\
\dot{v} & = & -\omega^2 \sin(\theta),\label{eq:Hunpert_v}
\end{eqnarray}
which is system (\ref{eq:Hpert_theta})--(\ref{eq:g1}) with
$\eps=0$ and $\beta=0$. The unperturbed system describes motions
of the pendulum with constant length and zero damping, so the
system has the first integral $H=const$. Unperturbed system
(\ref{eq:Hunpert_theta})--(\ref{eq:Hunpert_v}) has an oscillatory
solution if $H < \omega^2$ and a rotational solution if $H
> \omega^2$. If $H
= \omega^2$ the solution is a separatrix dividing oscillatory and
rotational domains in phase space ($\theta,\dot{\theta}$).

\subsection{Homoclinic bifurcations} In
order to apply Melnikov's criterion \cite{Melnikov,Guk} to a
homoclinic orbit we find the separatrix of the unperturbed system
(\ref{eq:Hunpert_theta})--(\ref{eq:Hunpert_v}) that goes through
the saddle-node point  $\theta = \pi$, $v = 0$. For this point we
have $H=\omega^2$ and with the use of (\ref{eq:H}) obtain $v^2 =
2\omega^2(1+\cos(\theta)) = 4\omega^2\cos^2(\theta/2)$. So, the
separatrix has the following form
\begin{equation}\label{eq:sprtrx}
    v = \dot{\theta} = \pm 2 \omega
    \cos\left(\frac{\theta}{2}\right).
\end{equation}
This equation allows for separation of variables and the following
integration
\begin{equation}\label{eq:sprtrx_int}
    \ln\left(\frac{1+\sin(\theta/2)}{\cos(\theta/2)}\right) = \pm
    \omega \left(\tau-\tau_0\right),
\end{equation}
where $\tau_0$ is a constant of integration. Potentiation of
(\ref{eq:sprtrx_int}) with some transformations yields
\begin{equation}\label{eq:sprtrx_cossin}
    \begin{array}{rcl}
    \displaystyle\cos\!\left(\frac{\theta}{2}\right) & = &
    \displaystyle\frac{1}{\cosh\left(\omega\left(\tau-\tau_0\right)\right)},\\[8pt]
    \displaystyle\sin\!\left(\frac{\theta}{2}\right) & = & \displaystyle\pm\tanh\left(\omega\left(\tau-\tau_0\right)\right).
    \end{array}
\end{equation}

Melnikov's distance between stable and unstable perturbed
separatrices is given by the following integral
\begin{eqnarray}\label{eq:Mpm}
    M^{\pm} & = & \int_{-\infty}^{\infty}\frac{\partial H}{\partial v}\,
    g_1(\theta(\tau),v(\tau),\tau)\d\tau,
\end{eqnarray}
where function $g_1$ is given in (\ref{eq:g1}) and $\frac{\partial
H}{\partial v} = v$.  With the use of (\ref{eq:sprtrx}) and
(\ref{eq:sprtrx_cossin}) integral (\ref{eq:Mpm}) has the following
expression (see Appendix $A$)
\begin{eqnarray}
    M^{\pm} & = &
  \frac{6\pi\eps\sin(\tau_0)}{\sinh\left(\frac{\pi}{2\omega}\right)}
  - 8\beta\omega^2.\label{eq:M}
\end{eqnarray}
This is a sign-changing function if the following inequality is
satisfied
\begin{equation}\label{eq:con}
    \frac{\eps}{\beta} >
    \frac{4\omega^2}{3\pi}\sinh\!\left(\frac{\pi}{2\omega}\right).
\end{equation}
 According to Melnikov's criterion \cite{Melnikov} when
condition (\ref{eq:con}) is satisfied then homoclinic structure
and complex dynamics appear near the separatrix. Condition
(\ref{eq:con}) means that complex dynamics appears only when the
amplitude of excitation is sufficiently large with respect to the
damping coefficient. The domain of possible complex dynamics
defined by inequality (\ref{eq:con}) is depicted in
Fig.~\ref{f:melnikov} and compared with numerical simulations in
Fig.~\ref{f:periods} for $\beta=0.05$. The minimum of the
right-hand side of (\ref{eq:con}) is reached at $\omega \approx
0.82$ and can be evaluated as $0.948$. Note that condition
(\ref{eq:con}) is similar to that of oscillator with quadratic
nonlinearity and external periodic excitation \cite{Kuznetsov}.
Similar inequality  for the pendulum with vertically vibrating
pivot was obtained in \cite{Koch_Leven}, where the right-hand side
function  is three times greater than that in (\ref{eq:con}).

\subsection{{Subharmonic bifurcations of
oscillatory orbits}} In order to apply Melnikov's criterion to
oscillatory orbits we find oscillatory solutions of the
unperturbed system (\ref{eq:Hunpert_theta})-(\ref{eq:Hunpert_v}).
For this reason we introduce the amplitude of oscillations $A$ so
that from the first integral for $v = 0$ we have $H = - \omega^2
\cos(A)$ and with the use of (\ref{eq:H}) we obtain $v^2 = 2
\omega^2\left(\cos(\theta) - \cos(A)\right) =
4\omega^2\left(\sin^2(A/2)-\sin^2(\theta/2)\right)$. Thus, we can
write the following equation
\begin{equation}\label{eq:oscl}
    v = \dot{\theta} = 2 \omega
    \sign(\dot{\theta})\sqrt{\sin^2\!\left(\frac{A}{2}\right)-\sin^2\!\left(\frac{\theta}{2}\right)},
\end{equation}
which allows for separation of variables. In order to integrate
(\ref{eq:oscl}) one usually introduces (see e.g. \cite{Tabor} or
\cite{Markeev1990}) a monotonically increasing phase $\psi$ such
that
\begin{equation}\label{eq:kpsi}\sin\left(\frac{\theta}{2}\right)=k
\sin\left(\psi\right),
\end{equation} where instead of amplitude $A$ one uses
$k = \sin\left(\frac{A}{2}\right) =
\sqrt{\frac{H+\omega^2}{2\omega^2}}$, which is called modulus in
elliptic integrals. We have from (\ref{eq:kpsi}) equation
(\ref{eq:oscl}) in the form $v = 2 \omega k \cos\psi$. This
equation along with time-differentiation of (\ref{eq:kpsi}) yields
$\omega\sqrt{1-k^2\sin^2\psi} = \dot\psi$. As a result of
integration starting from time $\tau_0$, when $\psi(\tau_0) = 0$,
we have $\psi = \am(\omega\left(\tau-\tau_0\right),k)$ and
consequently
\begin{equation}\label{eq:osc_v} v =
     2 \omega k \cn
     \left(\omega\left(\tau-\tau_0\right),k\right)
\end{equation}
     and
\begin{equation}\label{eq:osc_sincos}
    \begin{array}{rcl}
    \displaystyle \cos\!\left(\frac{\theta}{2}\right) & = & \dn\!\left(\omega\left(\tau-\tau_0\right),k\right),\\[8pt]
    \displaystyle \sin\!\left(\frac{\theta}{2}\right) & = & k\,\sn\!\left(\omega\left(\tau-\tau_0\right),k\right),
    \end{array}
\end{equation}
where $\am(\cdot,k)$, $\dn(\cdot,k)$, $\cn(\cdot,k)$, and
$\sn(\cdot,k)$ are the (elliptic) Jacobi functions. The elliptic
amplitude function $\am(\cdot,k)$ is the inverse of the incomplete
elliptic integral of the first kind $\int_0^{\psi}\frac{\d
\eta}{\sqrt{1-k^2\sin^2\eta}}$ and other elliptic functions are
defined as follows: $\sn(\cdot,k) = \sin\am(\cdot,k)$,
$\cn(\cdot,k) = \cos\am(\cdot,k)$, $\dn(\cdot,k) =
\sqrt{1-k^2\sn^2(\cdot,k)}$. The $k$ value follows from the
resonance condition stating that period of oscillation $\frac{4
K(k)}{\omega}$ and period of excitation $2\pi$ should be in
rational relation
\begin{equation}\label{eq:res}
    \frac{4 K(k)}{\omega}\,p =  2\pi\, q,
\end{equation}
where $p$ and $q$ are relatively prime natural numbers and $K(k)$
is the complete elliptic integral of the first kind. Thus, for
oscillation motion in resonance $p{:}q$ we have the following
subharmonic Melnikov's distance with the use of expressions
(\ref{eq:osc_v}), (\ref{eq:osc_sincos})
\begin{eqnarray}\label{eq:M_osc}
    M^{p/q} & = & \int_{0}^{2\pi q}\frac{\partial H}{\partial v}\,
    g_1(\theta(\tau),v(\tau),\tau)\d,
\end{eqnarray}
where function $g_1$ is given in (\ref{eq:g1}) and $\frac{\partial
H}{\partial v} = v$. Taking the integrals in Appendix~\ref{ap:2}
for $p=1$ and even $q=2,4,6,\ldots$ we have the following
Melnikov's distance
\begin{equation}\label{eq:M_osc1q}
M^{1/q} = \textstyle 4\omega^4 \!\left(\frac{3\eps
\pi\sin\tau_0}{\omega^2\sinh( K(k')/\omega)} - \scriptstyle
4\beta\left(E(k)-k'^2 K(k)\right)\right),
\end{equation}
where $k'^2=1-k^2$. This is a sign-changing function if the
following inequality is satisfied
\begin{equation}\label{eq:con2}
\frac{\eps}{\beta}
>\frac{4\omega^2}{3\pi}\left(E(k)-(k'^2)K(k)\right)\sinh\!\left(\frac{K(k')}{\omega}\right).
\end{equation}
\begin{figure*}[p]
\begin{minipage}[t]{0.5\textwidth}
\centering
\includegraphics[width=0.85\columnwidth]{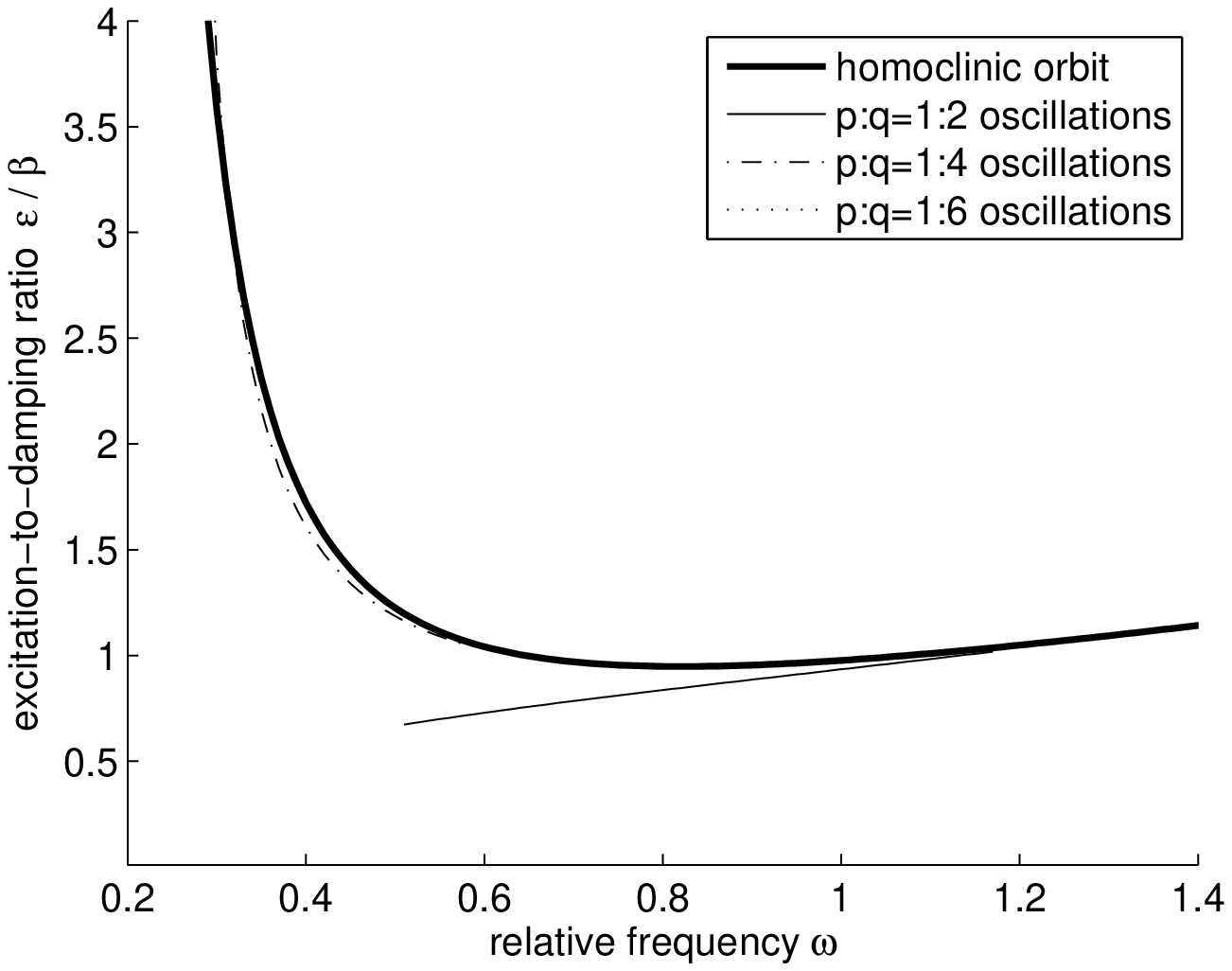}
\end{minipage}
\begin{minipage}[t]{0.5\textwidth}
\centering
\includegraphics[width=0.85\columnwidth]{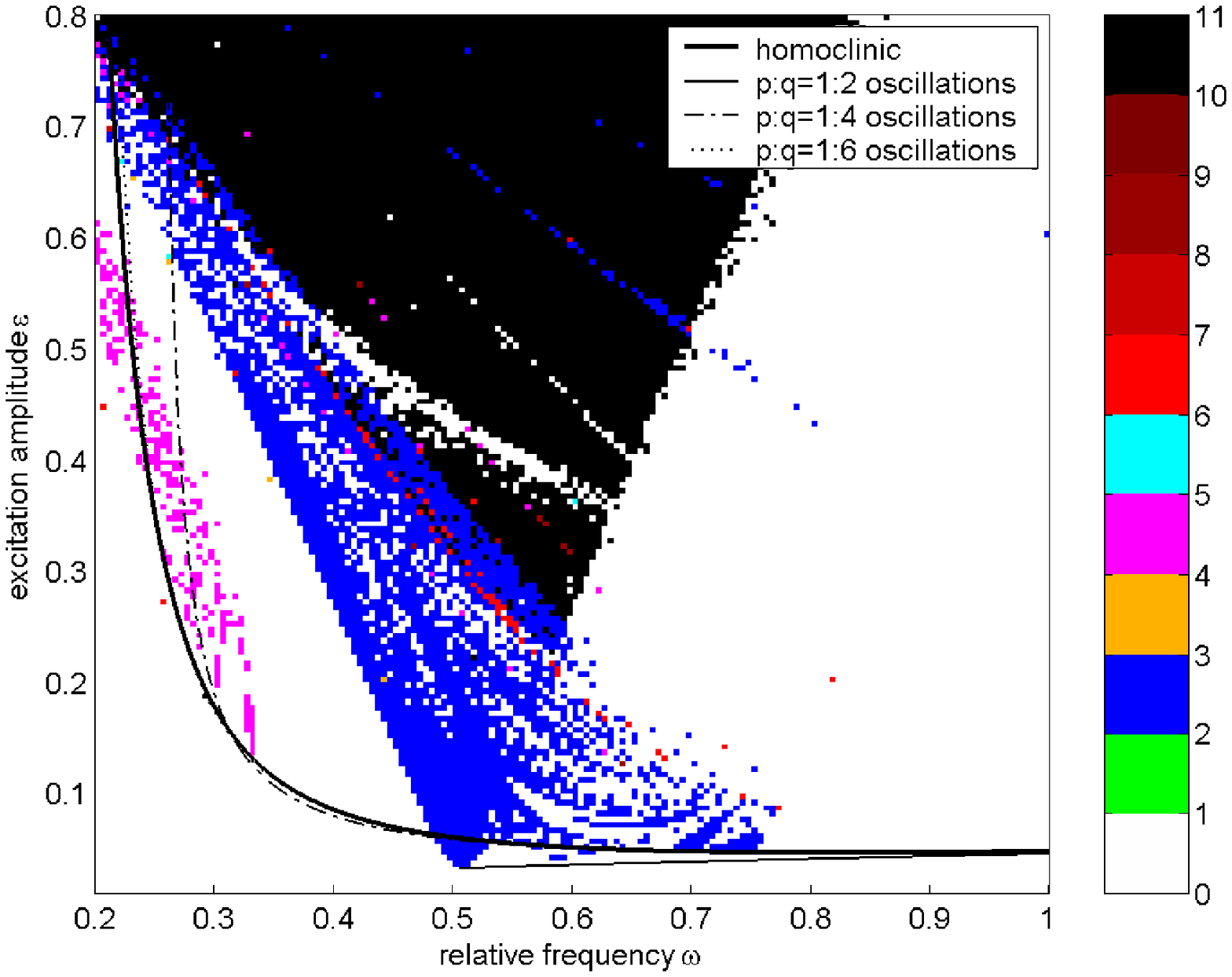}
\end{minipage}
\caption{Subharmonic bifurcation functions ($\eps/\beta$ -- left,
$\eps$ -- right) for $q=2$ (solid line), $q=4$ (dot-and-dash
line), $q=6$ (dotted line), and $p=1$ converge to the homoclinic
bifurcation function (bold solid line) that approximates the
domain of complex dynamics in (\ref{eq:con}). These functions
approximate the domains of corresponding resonant oscillations,
depicted (right) with blue ($q=2$), purple ($q=4$), and red
($q=6$) colors (see the color bar) on the parameter plane
$(\omega, \eps)$ at $\beta=0.05$. Chaotic regimes are shown with
black color.}\label{f:periods}\label{f:melnikov}
\end{figure*}
Condition (\ref{eq:con2}) means that corresponding $1{:}q$
resonant oscillations appear only when the amplitude of excitation
is sufficiently large with respect to the damping coefficient. The
domain of possible oscillations defined by inequality
(\ref{eq:con2}) is depicted in Fig.~\ref{f:melnikov} and compared
with numerical simulations 
for $\beta=0.05$. Note that condition (\ref{eq:con2}) is similar
to that of oscillator with quadratic nonlinearity and external
periodic excitation \cite{Kuznetsov}. Correspoding inequality for
the pendulum with vertically vibrating pivot, see
\cite{Koch_Leven}, has also three times greater right-hand side
function than that in (\ref{eq:con2}).

\subsection{Subharmonic bifurcations of rotational orbits}
In order to apply Melnikov's criterion to a rotational orbit we
find the solution of the unperturbed system
(\ref{eq:Hunpert_theta})--(\ref{eq:Hunpert_v}) for $H>\omega^2$.
Thus, with the use of (\ref{eq:H}) we obtain $\dot{\theta}=\pm
\sqrt{2}\sqrt{H+\omega^2\cos(\theta)} = \pm 2\omega
k\sqrt{1-\sin^2(\theta/2)/k^2}$, where "$\pm$" represents counter-
and clockwise rotations. Since the model is symmetric with respect
to the vertical axis, we will consider only the counterclockwise
rotation ("+" instead of "$\pm$"). This solution has the form
$\theta = 2\am(\omega k\left(\tau-\tau_0\right), 1/k)$, so we have
\begin{equation}\label{eq:rot_v}
    v = \dot{\theta} = 2 \omega k
    \dn\!\left(\omega k\left(\tau-\tau_0\right),
    \frac{1}{k}\right),
\end{equation}
\begin{equation}\label{eq:rot_cossin}
    \begin{array}{rcl}
    \displaystyle \cos\!\left(\frac{\theta}{2}\right) & = & \cn\!\left(\omega k\left(\tau-\tau_0\right), \frac{1}{k}\right),\\[8pt]
    \displaystyle \sin\!\left(\frac{\theta}{2}\right) & = & \sn\!\left(\omega k\left(\tau-\tau_0\right), \frac{1}{k}\right).
    \end{array}
\end{equation}
The value of $k$ follows from the resonance condition stating that
period of rotation $\frac{2 K(1/k)}{\omega k}$ and period of
excitation $2\pi$ should be in rational relation
\begin{equation}\label{eq:res}
    \frac{2 K\!\left(\frac{1}{k}\right)}{\omega k}\,r =  2\pi\,q,
\end{equation}
where $r$ and $q$ are relatively prime natural numbers. Thus, for
rotational motion in resonance $r{:}q$ we have the following
subharmonic Melnikov's distance
\begin{eqnarray}\label{eq:M_rot}
    M^{q/r} & = & \int_{0}^{2\pi q}\frac{\partial H}{\partial v}\,
    g_1(\theta(\tau),v(\tau),\tau)\d\tau,
\end{eqnarray}
where function $g_1$ is given in (\ref{eq:g1}) and $\frac{\partial
H}{\partial v} = v$. Taking the integrals in Appendix~\ref{ap:3}
for $r=1$ and $q=1,2,3,\ldots$ we have the following Melnikov's
distance
\begin{eqnarray}\label{eq:M_rotq1}
M^{q/1} & = & -2\omega^2 \left(\frac{3\eps
\pi}{\omega^2}\frac{\sin\tau_0}{\sinh\!\left(\frac{K'}{\omega
k}\right)} - 4\beta k E\!\left(\frac{1}{k}\right)\right),
\end{eqnarray}
where $K' = K\!\left(\sqrt{1-1/k^2}\right)$. This is a
sign-changing function if the following inequality is satisfied
\begin{equation}\label{eq:con3}
\frac{\eps}{\beta}
>\frac{4\omega^2 k}{3\pi}\,E\!\left(\frac{1}{k}\right)\sinh\!\left(\frac{K'}{\omega
k}\right).
\end{equation}

\begin{figure*}[p]
\begin{minipage}[t]{0.5\textwidth}
\centering
\includegraphics[width=0.85\columnwidth]{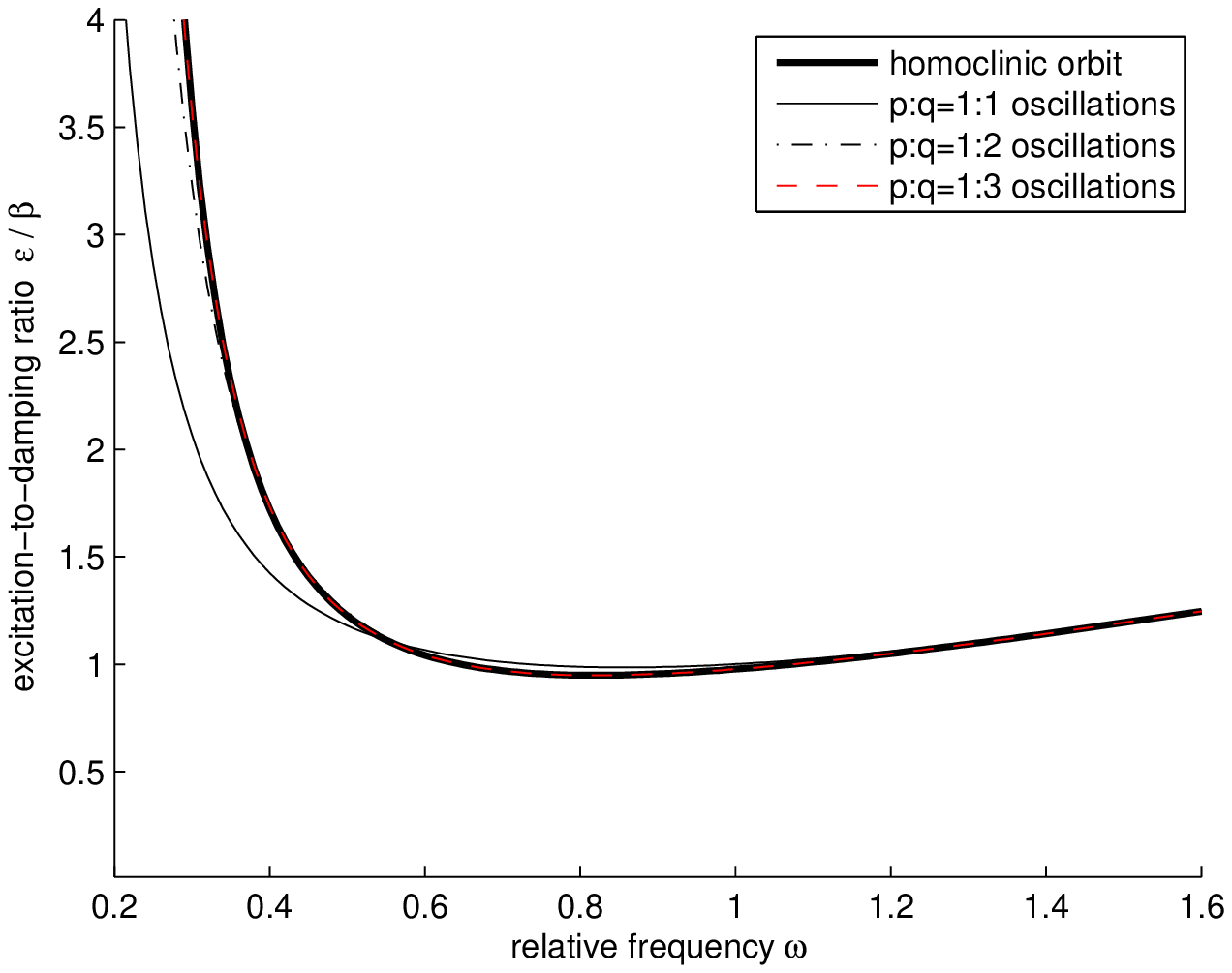}
\end{minipage}
\begin{minipage}[t]{0.5\textwidth}
\centering
\includegraphics[width=0.85\columnwidth]{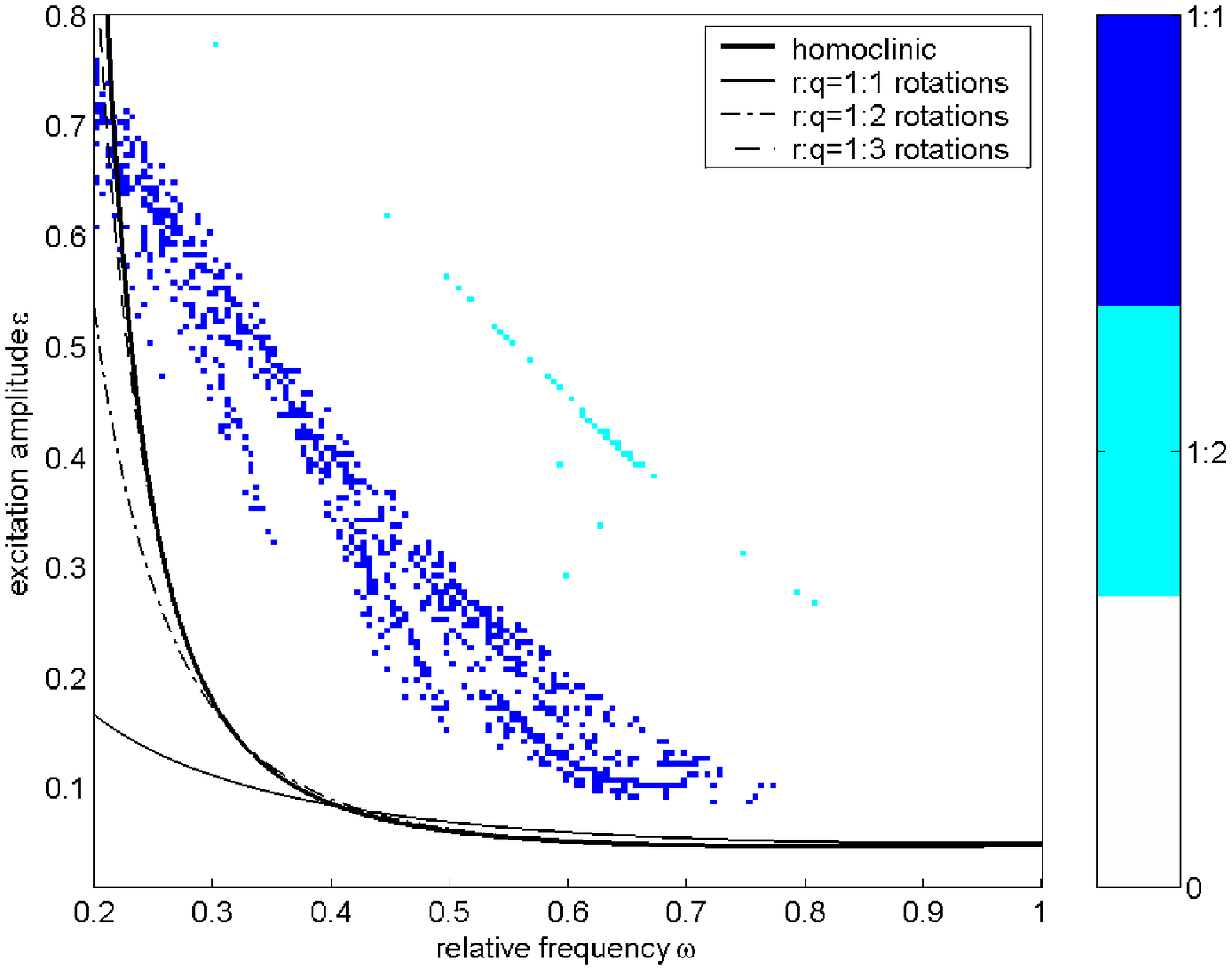}
\end{minipage}
\caption{Subharmonic bifurcation functions ($\eps/\beta$ -- left,
$\eps$ -- right), which are the boundaries in (\ref{eq:con3}) for
$q=1$ (solid line), $q=2$ (dot-and-dash line), $q=3$ (dashed
line), and $r=1$, converge to the homoclinic bifurcation function
that approximates the domain of complex dynamics in
(\ref{eq:con}). These functions approximate the domains of
corresponding resonant rotations, depicted with blue ($q=1$) and
azure ($q=2$) colors (see the color bar) on the parameter plane
$(\omega, \eps)$ at $\beta=0.05$.}
\label{f:melnikov_rot}\label{f:periods_rot}
\end{figure*}

Condition (\ref{eq:con3}) means that corresponding $1{:}q$
resonant rotations appear only when the amplitude of excitation is
sufficiently large with respect to the damping coefficient. The
domain of possible rotations defined by inequality (\ref{eq:con3})
is depicted in Fig.~\ref{f:melnikov_rot} and compared with
numerical simulations 
for $\beta=0.05$. Although numerically found rotations $1{:}2$ are
not monotone. These are rotation-oscillation regimes. Note that
condition (\ref{eq:con3}) is similar to that of oscillator with
quadratic nonlinearity and external periodic excitation
\cite{Kuznetsov}. For the pendulum with vertically vibrating pivot
similar inequality was obtained in \cite{Koch_Leven}, with the
right-hand side function being three times greater than that in
(\ref{eq:con3}). Thus, the Melnikov approach leads to similar
results for different models of pendula.

\section{{Averaging method: Superharmonic bifurcations of rotational orbits}}\label{s:avg} We
study resonant rotations
\begin{equation}\label{eq:resRQ}
r{:}q = 1{:}1,\, 2{:}1,\, 3{:}1,\ldots ,
\end{equation}
where $r$ is the number of full rotations during $q$ complete
periods of excitation. The new assumption here is that the
relative eigenfrequency $\omega \ll 1$ is small of the same order
with damping parameter $\beta \sim \omega \ll 1$, e.g. because of
the small gravitation $g$ or high excitation frequency $\Omega \gg
1$, while the excitation amplitude $\eps$ is not small. Then,
unperturbed system (equation (\ref{eq:swing2}) with $\omega = 0$)
has the specific angular momentum or sector velocity $s =
\left(1+\eps\varphi(\tau)\right)^2\dot{\theta}$ as its first
integral. The unperturbed system is not Hamiltonian, so we cannot
apply Melnikov's analysis to obtain domains of existence of
corresponding rotations. For this purpose we use the method of
averaging, \cite{BogMit,Vol_Morg}. Equation (\ref{eq:swing2}) can
be written in the form of the system
\begin{eqnarray}
\dot{\theta} & = & \frac{s}{\left(1+\eps\varphi(\tau)\right)^2},\label{eq:thetadot}\\
\dot{s} & = & \omega^2 f(\theta,s,\tau),\label{eq:sdot}
\end{eqnarray}
where the perturbation function is the following
\begin{equation}\label{eq:f1}
f(\theta,s,\tau) \equiv -\frac{\beta}{\omega}\, s -
\left(1+\eps\varphi(\tau)\right) \sin(\theta),
\end{equation}
with the ratio $\beta/\omega = O(1)$ that can be considered as a new
parameter. The unperturbed system 
has the following solution
\begin{equation}\label{eq:solform}
\theta_0 = s_0\,\Phi(\tau)+\vartheta_0,
\end{equation}
where $\vartheta_0$ is the constant phase shift, $s_0$ is the
constant sector velocity, and $\Phi(\tau)$ denotes the following
integral
\begin{equation}\label{eq:Phi}
\Phi(\tau) =\int_0^{\tau}\frac{\d
\eta}{\left(1+\eps\varphi(\eta)\right)^2}.
\end{equation}
We choose such constants $s_0$ and $\vartheta_0$, that they
approximate the perturbed solution, i.e.
 $\theta = \theta_0 + o(1)$. In order to do so
we take a resonance condition from (\ref{eq:resRQ}) along with the
following \emph{averaged equation} of (\ref{eq:sdot})
\begin{equation}\label{eq:sdotAVG} \dot{\bar{s}} =  \omega^2
F(\bar{s}),
\end{equation}
where the first order approximation function $F$ is derived via
the substitution in $f$ variable $\theta$ by the expression
$\bar{s}\,\Phi(\tau) +\vartheta$ and taking time-average of $f$ as
if the corresponding \emph{averaged variable} $\bar{s}$ and
$\vartheta$ are constant, see \cite{Vol_Morg}:
\begin{eqnarray*}
  F(\bar{s}) & = & \frac{1}{2\pi q r}\int\limits_0^{2\pi q r}
f(\bar{s}\,\Phi(\tau)+\vartheta,\bar{s},\tau)\d \tau\\
& = & -\frac{\beta}{\omega}\, \bar{s} - \int\limits_0^{2\pi q
r}\!\frac{1+\eps\varphi(\tau)}{2\pi q r}
\sin(\bar{s}\,\Phi(\tau)+\vartheta) \d \tau\\
 &=& -\frac{\beta}{\omega}\, \bar{s} - A(\bar{s}) \cos\!\left(\vartheta\right)- B(\bar{s})
\sin\!\left(\vartheta\right),
\end{eqnarray*}
where $A$ and $B$ denote the following integrals
\begin{eqnarray}
A(\bar{s}) & = & \int\limits_0^{2\pi r
q}\!\frac{1+\eps\varphi(\tau)}{2\pi r q}
\sin\!\left(\bar{s}\,\Phi(\tau)\right)\!\d \tau,\label{eq:A}\\
B(\bar{s}) & = & \int\limits_0^{2\pi r
q}\!\frac{1+\eps\varphi(\tau)}{2\pi r q}
\cos\!\left(\bar{s}\,\Phi(\tau)\right)\!\d \tau.\label{eq:B}
\end{eqnarray}
The period of averaging is chosen $2\pi r q$ to contain integer
numbers of motion and excitation periods. Notice that the averaged equation of (\ref{eq:thetadot})
$
  \dot{\bar{\theta}} = \bar{s}/\left(1+\eps\varphi(\tau)\right)^2
$
does not influence the dynamics of $\bar{s}$ in (\ref{eq:sdotAVG}) and for constant (steady state) $\bar{s}$ has the solution $\bar{\theta} = \bar{s}\,\Phi(\tau) +\vartheta$.

The steady state value of $\bar{s}$ follows from resonance
condition (\ref{eq:resRQ}) written with the use of solution
(\ref{eq:solform}) and stating that period of rotation and period
of excitation should be in rational relation
\begin{equation}\label{eq:resR}
2\pi \,r = s_0\,q\,\Phi(2\pi),
\end{equation}
where $r$ and $q$ are relatively prime natural numbers from
(\ref{eq:resRQ}). We have from (\ref{eq:resR}) the approximate
steady state value, $\bar{s} = s_0$, where
\begin{equation}\label{eq:sbar}
    s_0 = \frac{r}{q}\,\frac{2\pi}{\Phi(2\pi)}.
\end{equation}
Values of $\vartheta$ we find from the averaged equation
(\ref{eq:sdotAVG}) when we set $\dot{\bar{s}}=0$ and substitute
$\bar{s}$ by its steady state value $s_0$ expressed in
(\ref{eq:sbar}), so that $F(s_0) = 0$:
\begin{equation}
A(s_0) \cos\!\left(\vartheta_0\right)+B(s_0)
\sin\!\left(\vartheta_0\right) = -\frac{\beta}{\omega}\,s_0
\label{eq:SS}
\end{equation}
Thus, we find $\vartheta = \vartheta_0$, that takes values form
two branches ($\pm$) of the solution
\begin{equation}\label{eq:thetabar}
    \vartheta_0 = \vartheta^* + \pi \pm
    \arccos\!\left(\frac{s_0}{\sqrt{A^2(s_0)+B^2(s_0)}}\,\frac{\beta}{\omega}\right),
\end{equation}
where the constant $\vartheta^*$ can be expressed as follows
\begin{equation*}
\vartheta^* =
\sign\!\left(B(s_0)\right)\arccos\!\left(\frac{A(s_0)}{\sqrt{A^2(s_0)+B^2(s_0)}}\right)+2\pi
n,
\end{equation*}
with $n$ being an integer number. Thus, the domain of existence of
corresponding regular rotations is approximated by the following
condition that equation (\ref{eq:SS}) has the solution expressed
in (\ref{eq:thetabar}),
\begin{equation}\label{eq:con4}
\frac{\omega}{\beta} \geq \frac{s_0}{\sqrt{A^2(s_0)+B^2(s_0)}}.
\end{equation}

The averaged variables $\bar{s}$ and $\bar{\theta}$ approximate the slow variable $s =
\bar{s} + o(1)$ and fast phase $\theta = \bar{\theta} + o(1)$,
which solve system (\ref{eq:thetadot})--(\ref{eq:sdot}). The
solution for regular rotational motion in resonance $r{:}q$ can
be written as follows
\begin{equation}\label{eq:solform2}
\theta = \bar{s}\,\Phi(\tau) +\vartheta + o(1) =
s_0\,\Phi(\tau,\eps) + \vartheta_0  + o(1),
\end{equation}
where $s_0$ and $\vartheta_0$ are defined in
(\ref{eq:sbar}) and (\ref{eq:thetabar}).

We obtain here only the first order approximation with the use of
averaged equation (\ref{eq:sdotAVG}). In order to obtain higher
order approximations the general averaging scheme by Volosov can
be applied to system (\ref{eq:thetadot})-(\ref{eq:sdot}), see
\cite{Vol_Morg}.

\subsection{Stability analysis}
\begin{figure*}[p]
\begin{minipage}[t]{0.5\textwidth}
\centering
\includegraphics[width=0.8\columnwidth]{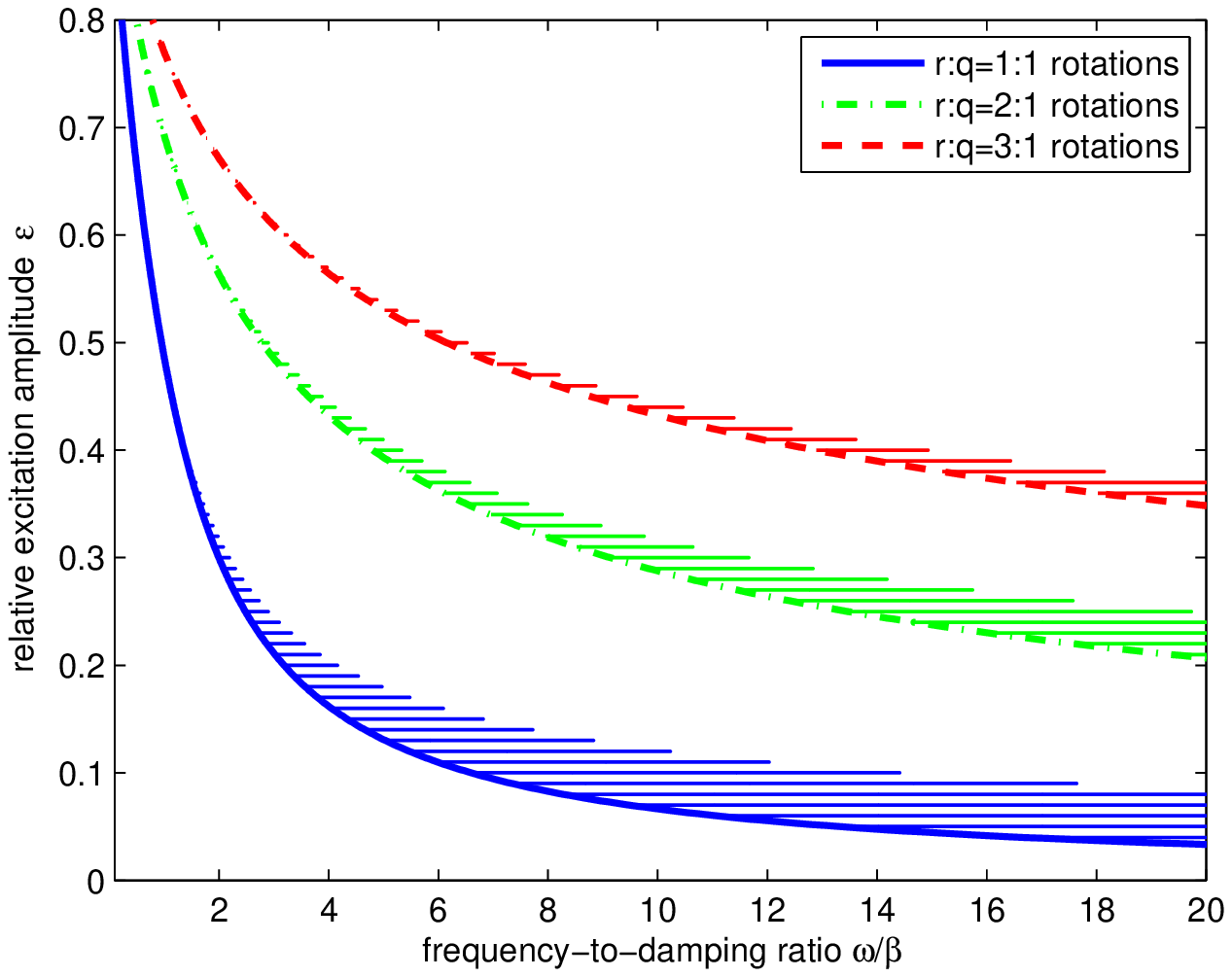}
\end{minipage}
\begin{minipage}[t]{0.5\textwidth}
\centering
\includegraphics[width=0.85\columnwidth]{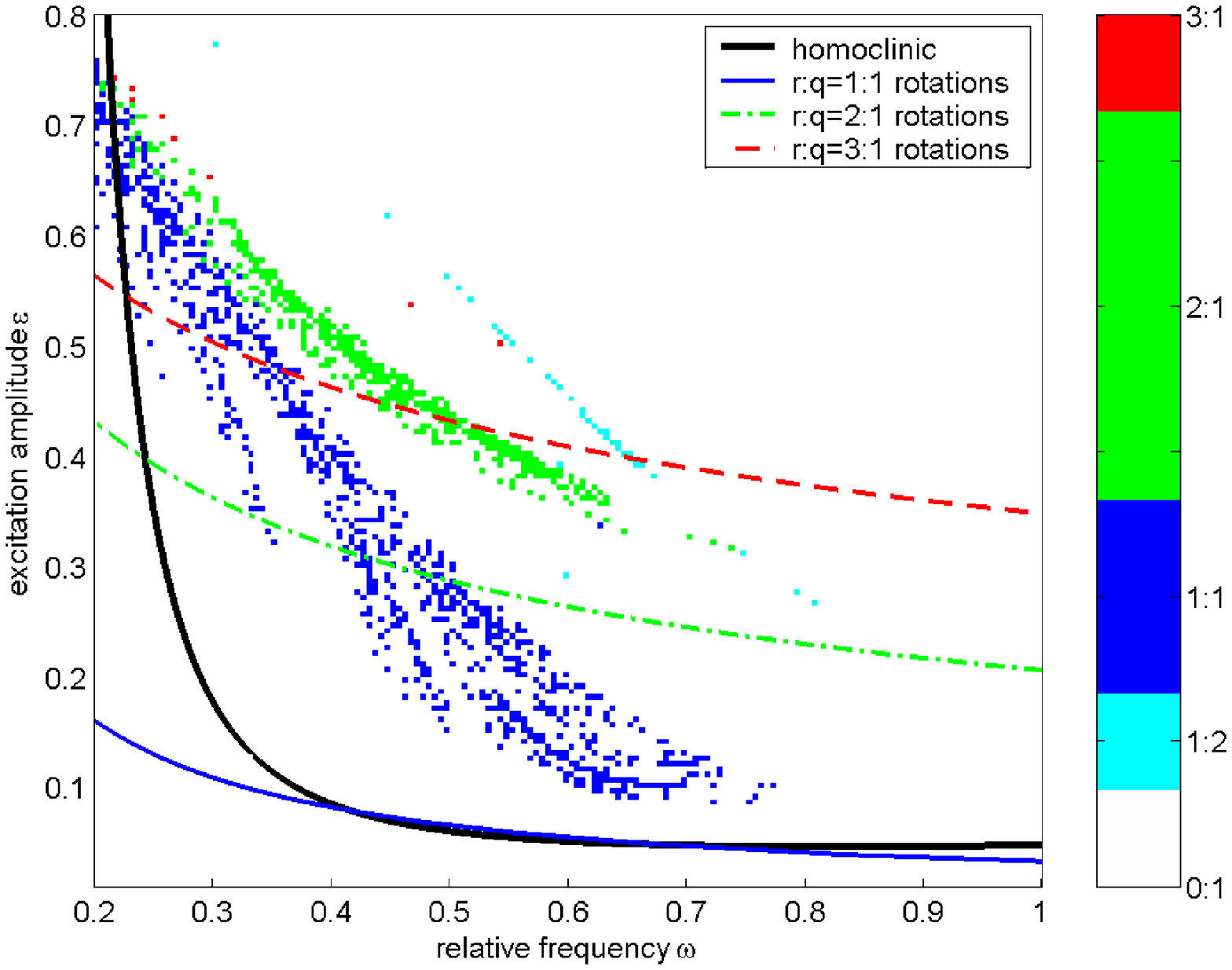}
\end{minipage}
\caption{Superharmonic bifurcation functions which are the
boundaries in (\ref{eq:con4}) for $r=1$, $r=2$, $r=3$, and $q=1$
approximate domains of corresponding rotations, depicted (right)
with blue ($r=1$), green ($r=2$), and red ($r=3$) colors (see the
color bar) on the parameter plane $(\omega, \eps)$ at
$\beta=0.05$. Hatch lines (left) denote the region where
corresponding solution exists but it is unstable.}
\label{f:super_rot}\label{f:super_rot_data}
\end{figure*}

In order to study the stability of a solution $\bar s$ we perturb
it by small value $\eta$ in (\ref{eq:sdotAVG}). Thus we have the
linearized equation
\begin{equation}\label{eq:sdotAVGlin} \dot{\eta} =  \omega^2
F^{\prime}(\bar{s})\eta,
\end{equation}
with the derivative of function $F$ in the form
\begin{eqnarray*}
  F^{\prime}(\bar{s}) & = & -\frac{\beta}{\omega} - A^{\prime}(\bar{s}) \cos\!\left(\vartheta\right)
    - B^{\prime}(\bar{s})\sin\!\left(\vartheta\right),
\end{eqnarray*}
where $A^{\prime}$ and $B^{\prime}$ are the derivatives of
integrals $A$ and $B$ in (\ref{eq:A}) and (\ref{eq:B}):
\begin{eqnarray}
A^{\prime}(\bar{s}) & = & \int\limits_0^{2\pi r
q}\!\frac{1+\eps\varphi(\tau)}{2\pi r q}\,
\Phi(\tau)\cos\!\left(\bar{s}\,\Phi(\tau)\right)\!\d \tau,\\
B^{\prime}(\bar{s}) & = & -\int\limits_0^{2\pi r
q}\!\frac{1+\eps\varphi(\tau)}{2\pi r q}\,
\Phi(\tau)\sin\!\left(\bar{s}\,\Phi(\tau)\right)\!\d \tau.
\end{eqnarray}
According to Lyapunov's theorem on stability based on a linear
approximation the instability or asymptotic stability of the
solution $\bar{s}$ of equation (\ref{eq:sdotAVG}) is determined by
the instability or asymptotic stability of the linearized equation
(\ref{eq:sdotAVGlin}). Thus, for steady state solution, $\bar{s} =
s_0$, we have the condition of asymptotic stability
$F^{\prime}(s_0) < 0$:
\begin{equation}
    A^{\prime}(s_0) \cos\!\left(\vartheta_0\right) + B^{\prime}(s_0)\sin\!\left(\vartheta_0\right)
    > -\frac{\beta}{\omega}.
\end{equation}
This condition is checked numerically for various parameters
$\eps$ and $\omega/\beta$, see Fig.~\ref{f:super_rot} (left). It
turns out that only one branch of the solution, that corresponds
to plus ($+$) in (\ref{eq:thetabar}), can be stable. Moreover
there are regions (denoted with hatch lines in
Fig.~\ref{f:super_rot}) where both existing branches of the
solution are unstable. Superharmonic bifurcations of rotational orbits
happen on the upper borders of the corresponding hatched areas.
Although it would be natural to expect that exact (not approximate)
borders of existence domain coincide with the border of stability domain.

\subsection{Comparison with direct simulations}
In order to find the boundaries for rotational regimes with relative angular velocities $r=1,2,3$
the first approximation is enough, because excitation amplitude
$\eps$ is not small in contrast to the quasi-linear approach in
\cite{BelSey2012}, where both $\eps$ and $\omega$ are assumed to
be small and higher order approximations of the averaging method
are needed to obtain similar boundaries. Notice that it is also
possible to consider not small damping $\beta$ when sector
velocity $s$ is small, because they are multiplied in the first
term of perturbation function (\ref{eq:f1}).

The right hand side in condition (\ref{eq:con4}) can be calculated for any particular $\eps$, $r$, and $q$, see Fig.~\ref{f:super_rot}, where $\varphi(\tau)=\cos(\tau)$. In order to do that, first we calculate numerically (\ref{eq:Phi}) as function of $\tau$, then we take integrals (\ref{eq:A}) and (\ref{eq:B}) for $s_0$ obtained from (\ref{eq:sbar}). In right Fig.~\ref{f:super_rot} we depict the points where numerical simulation converged to regular rotations. The points have different colors for relative angular velocities $r=1,2,3$. All these points are above the corresponding boundaries of existence domains which have same colors. Thus, existence condition  (\ref{eq:con4}) is satisfied for all numerically obtained rotational solutions.

\section{Conclusion}
For the pendulum with variable length we derived analytical
formulas for the boundaries of bifurcations in the space of three
parameters: the relative frequency of excitation, amplitude of
excitation, and damping. The boundaries for homoclinic bifurcation
separating the domain of only stationary and oscillatory regimes
from the domain of more complex dynamics, subharmonic oscillations
and subharmonic rotations are obtained using Melnikov's method
under assumption of small damping and excitation amplitude. For
the analysis of superharmonic bifurcations of rotational orbits
the method of averaging is used assuming smallness of relative
excitation frequency rather than that of excitation amplitude.
Both methods allow to obtain in the first approximation the basic
rotational orbit with angular velocity equal to the excitation
frequency, $1{:}1$. Small excitation frequency (or small gravity)
allows to introduce the unperturbed system with the conservation
of angular momentum, so that faster rotations are found in the
first approximation by the method of averaging.  In
Figs.~\ref{f:periods}, \ref{f:periods_rot}, and
\ref{f:super_rot_data} it is shown that the boundaries for complex
dynamics, subharmonic oscillations, and superharmonic rotations
are in good agreement with the results of numerical simulation.

\begin{acknowledgements}
This research was partly supported by the Austrian Science Fund
(FWF) under Grant P25979-N25 and by the Russian Foundation for
Basic Research (RFBR), Grant No. 13-01-00261.
\end{acknowledgements}

\section*{Appendices}
\appendix
\section{Melnikov function for homoclinic orbit}\label{ap:1}\vspace{-10pt}
\begin{eqnarray}
    M^{\pm} & = &
  8\eps\omega^2\int_{-\infty}^{\infty}\frac{\sin(\tau)\,\d\tau}{\cosh^2\!\left(\omega\left(\tau-\tau_0\right)\right)}\nonumber\\
  & &-\,4\beta\omega^3\int_{-\infty}^{\infty}\frac{\d\tau}{\cosh^2\!\left(\omega\left(\tau-\tau_0\right)\right)}\nonumber\\
  & & +\,  4\eps\omega^3\int_{-\infty}^{\infty}\frac{\cos\!\left(\tau\right)\sinh\!\left(\omega\left(\tau-\tau_0\right)\right)}{\cosh^3\!
  \left(\omega\left(\tau-\tau_0\right)\right)}\d\tau
\end{eqnarray}
we denote as $M^{\pm} = 8\eps\omega^2 I_1 - 4\beta\omega^3 I_2 +
4\eps\omega^3 I_3$, where
\begin{eqnarray}
I_{1} & = & \int_{-\infty}^{\infty}\frac{\sin(\tau)
\d\tau}{\cosh^2\!\left(\omega \left(\tau-\tau_0\right)\right)}
 = \frac{1}{\omega}\!\int_{-\infty}^{\infty}\frac{\sin\!\left(\tau_0+\eta/\omega\right) \d\eta}{\cosh^2\!\left(\eta\right)} \nonumber\\
& = &
\frac{\sin\!\left(\tau_0\right)}{\omega}\!\int_{-\infty}^{\infty}\!\frac{\cos\!\left(\eta/\omega\right)
\!\d\eta}{\cosh^2\!\left(\eta\right)}+
\frac{\cos\!\left(\tau_0\right)}{\omega}\!\int_{-\infty}^{\infty}\!\frac{\sin\!\left(\eta/\omega\right)
\!\d\eta}{\cosh^2\!\left(\eta\right)}.
\end{eqnarray}
The integral
$\int_{-\infty}^{\infty}\frac{\sin\!\left(\eta/\omega\right)
\d\eta}{\cosh^2\!\left(\eta\right)}$ is zero because its integrand
is an odd function, while the other integral has even integrand
and can be calculated as follows $
\int_{-\infty}^{\infty}\frac{\cos(t/\omega)}{\cosh^2 t}\d t =
\frac{\pi}{\omega\sinh(\pi/2\omega)}$, hence, the first term has
the expression
\begin{eqnarray}
I_{1} & = &
\frac{\pi\sin\left(\tau_0\right)}{\omega^2\sinh(\pi/2\omega)}.
\label{eq:I1}
\end{eqnarray}
The second integral can be calculated as follows
\begin{eqnarray}
I_2 & = &
\int_{-\infty}^{\infty}\frac{\d\tau}{\cosh^2\!\left(\omega
\left(\tau-\tau_0\right)\right)}
 = \frac{1}{\omega}\int_{-\infty}^{\infty}\frac{\d s}{\cosh^2\!\left(s\right)} = \frac{2}{\omega}, \label{eq:I2}
\end{eqnarray}
while the integral $I_3$ can be converted to $I_1$ via integration
by parts using the relation $\frac{\sinh(s)\d s}{\cosh^3(s)} =
-\frac{1}{2}\d\frac{1}{\cosh^2(s)}$ as
\begin{eqnarray}
I_3 & = & \int_{-\infty}^{\infty}\frac{\cos(\tau)\,\sinh\left(\omega\left(\tau-\tau_0\right)\right)}{\cosh^3\!\left(\omega\left(\tau-\tau_0\right)\right)}\d\tau,\nonumber\\
& = & \frac{1}{\omega}\int_{-\infty}^{\infty}\frac{\cos(\tau_0 +
\eta/\omega)\,\sinh\left(\eta\right)}{\cosh^3\!\left(\eta\right)}\d
\eta\nonumber\\
& = & -\frac{1}{2\omega^2}\left.\frac{\cos(\tau_0 +
\eta/\omega)}{\cosh^2\!\left(\eta\right)}\right|_{-\infty}^{\infty}
- \frac{1}{2\omega^2}\int_{-\infty}^{\infty}\frac{\sin(\tau_0 +
\eta/\omega)}{\cosh^2\!\left(\eta\right)}\d\eta\nonumber\\
& = & - \frac{I_1}{2\omega}. \label{eq:I3}
\end{eqnarray}
Thus, $M^{\pm} = 8\eps\omega^2 I_1 - 4\beta\omega^3 I_2 +
4\eps\omega^3 I_3 =
\frac{6\pi\eps\sin\left(\tau_0\right)}{\sinh(\pi/2\omega)} -
8\beta\omega^2$.

\section{Melnikov function for subharmonic oscillations}\label{ap:2}\vspace{-10pt}
\begin{eqnarray}
    M^{p/q} & = & 4\omega^2 k^2 \int_{0}^{2\pi q}\left(2\eps\sin(\tau)-\beta\omega\right)\cn^2\!\left(\omega\left(\tau-\tau_0\right),k\right)\d\tau\nonumber\\
  & & +\,4\eps \omega^3 k^2 \int_{0}^{2\pi
  q}\cos(\tau)\sn\!\left(\omega\left(\tau-\tau_0\right),k\right)\nonumber\\
  &&\times\dn\!\left(\omega\left(\tau-\tau_0\right),k\right)\cn\!\left(\omega\left(\tau-\tau_0\right),k\right)\d\tau,
\end{eqnarray}
so we denote $M^{p/q} = 8\eps\omega^2 k^2 I_1 - 4\beta\omega^3 k^2
I_2 + 4\eps\omega^3 k^2 I_3$,
\begin{eqnarray}
I_1 & = & \int_{0}^{2\pi q}\sin(\tau)\cn^2\!\left(\omega\left(\tau-\tau_0\right),k\right)\d\tau\nonumber\\
& = & -\left.
\cos(\tau)\cn^2\!\left(\omega\left(\tau-\tau_0\right),k\right)\right|_{0}^{2\pi
q}  - 2\omega I_3 = - 2\omega I_3,\\
I_2 & = & \int_{0}^{2\pi q}\cn^2\!\left(\omega\left(\tau-\tau_0\right),k\right)\d\tau,\\
I_3 & = & \int_{0}^{2\pi
q}\cos(\tau)\cn\!\left(\omega\left(\tau-\tau_0\right),k\right)\sn\!\left(\omega\left(\tau-\tau_0\right),k\right)\nonumber\\
  &&\times\dn\!\left(\omega\left(\tau-\tau_0\right),k\right)\d\tau,
\end{eqnarray}
where we use the formula $\frac{\d \cn(u)}{\d u} = -
\sn(u)\,\dn(u)$. Thus, we have $M^{q/p} = -12\eps\omega^3 k^2 I_3
- 4\beta\omega^3 k^2 I_2$.

\begin{eqnarray}
I_1 & = & \int_{0}^{2\pi q}\sin(\tau)\cn^2\!\left(\omega\left(\tau-\tau_0\right),k\right)\d\tau\nonumber\\
& = & \frac{1}{\omega}\int_{0}^{2\pi q \omega}\sin(\tau_0 +
s/\omega)\cn^2\!\left(s,k\right)\d s\nonumber\\
& = & \frac{\sin(\tau_0)}{\omega}\int_{0}^{2\pi q
\omega}\cos(s/\omega)\cn^2\!\left(s,k\right)\d s
\end{eqnarray}

The integral $I_3$ vanishes except for $p=1$ and even $q$. In this
case we have from (312.02) in \cite{Byrd_Friedman}
$$
I_2 = \frac{4\omega}{k^2}\left(E(k)-(k'^2)K(k)\right), \quad I_3 =
-\frac{\pi}{k^2\omega}\frac{\sin\tau_0}{\sinh(K(k'))},
$$
where $k'^2=1-k^2$. So we have (\ref{eq:M_osc1q}).

\section{Melnikov function for subharmonic rotations}\label{ap:3}
\vspace{-10pt}
\begin{eqnarray}
    M^{q/r} & = & 4\omega^2 k^2 \int_{0}^{2\pi q}\left(2\eps\sin(\tau)-\beta\omega\right)\dn^2\!\left(\omega k \left(\tau-\tau_0\right),\frac{1}{k}\right)\d\tau\nonumber\\
  & & +\,4\eps \omega^3 k \int_{0}^{2\pi
  q}\cos(\tau)\sn\!\left(\omega k\left(\tau-\tau_0\right),\frac{1}{k}\right)\nonumber\\
  &&\times\dn\!\left(\omega k\left(\tau-\tau_0\right),\frac{1}{k}\right)\cn\!\left(\omega
  k\left(\tau-\tau_0\right),\frac{1}{k}\right)\d\tau,
\end{eqnarray}
we denote $M^{q/r} = 8\eps\omega^2 k^2 I_1 - 4\beta\omega^3 k^2
I_2 + 4\eps\omega^3 k I_3$,
\begin{eqnarray}
I_1 & = & \int_{0}^{2\pi q}\sin(\tau)\dn^2\!\left(\omega k\left(\tau-\tau_0\right),\frac{1}{k}\right)\d\tau = - \frac{2\omega}{k} I_3\nonumber\\
& & -\left. \cos(\tau)\,\dn^2\!\left(\omega
k\left(\tau-\tau_0\right),\frac{1}{k}\right)\right|_{0}^{2\pi
q}   = - \frac{2\omega}{k} I_3,\\
I_2 & = & \int_{0}^{2\pi q}\dn^2\!\left(\omega k\left(\tau-\tau_0\right),\frac{1}{k}\right)\d\tau,\\
I_3 & = &  \int_{0}^{2\pi
  q}\cos\!\left(\tau\right)\sn\!\left(\omega k\left(\tau-\tau_0\right),
  \frac{1}{k}\right)\nonumber\\
  &&\times\dn\!\left(\omega k\left(\tau-\tau_0\right),\frac{1}{k}\right)\cn\! \left(\omega k\left(\tau-\tau_0\right),\frac{1}{k}\right)\d\tau,
\end{eqnarray}
where we use the formula $\frac{\d \dn(u,\frac{1}{k})}{\d u} = -
\frac{\sn(u,\frac{1}{k})\,\cn(u,\frac{1}{k})}{k^2}$. Thus, we have
$M^{q/r} = -12\eps\omega^3 k^2 I_3 - 4\beta\omega^3 k^2 I_2$.

The integral $I_3$ vanishes except for $r=1$. In this case we have
$$
I_2 = \frac{2}{\omega k}\, E\!\left(\frac{1}{k}\right), \quad I_3
= -\frac{\pi}{k^2\omega}\frac{\sin\tau_0}{\sinh(K')},
$$
where $K' = K\!\left(\sqrt{1-1/k^2}\right)$. So we have
(\ref{eq:M_rotq1})
.



\end{document}